\documentclass[twocolumn,english,prl,showpacs]{revtex4-1}
\usepackage[colorlinks=true,urlcolor=blue,citecolor=blue,linkcolor=blue]{hyperref}
\usepackage[sort&compress]{natbib}
\usepackage[T1]{fontenc}
\usepackage[latin9]{inputenc}
\usepackage{amssymb}
\usepackage{graphicx}
\usepackage{amsmath,color}
\usepackage{mathrsfs}
\usepackage{float}
\usepackage{indentfirst}
\usepackage{babel}
\usepackage{color}

\makeatletter

\begin{document}

\title{Origin of the unusual strong suppression of low frequency antiferromagnetic fluctuations in underdoped HgBa$_2$CuO$_{4+\delta}$}

\author{Jia-Wei Mei$^{1}$}
\author{Alexey A. Soluyanov$^{1}$}
\author{T. M. Rice$^{1,2}$}
\affiliation{${^1}$Institute for Theoretical Physics, ETH Z\"urich, 8093 Z\"urich, Switzerland}
\affiliation{${^2}$Brookhaven National Laboratories, Upton, New York, USA}

\date{\today}

\begin{abstract}
Generally strong charge and magnetic inhomogeneities are observed in NQR/NMR experiments on underdoped cuprates. It is not the case for the underdoped HgBa$_2$CuO$_{4+\delta}$, the most symmetric and highest T$_c$ single layer cuprate, whose magnetic inhomogeneity is strongly suppressed. Also neutron scattering experiments reveal a unique pair of weakly dispersive magnetic modes in this material. We propose that these special properties stem from the symmetric positioning of the O-dopants between adjacent CuO$_2$ layers that lead to a strong superexchange interaction between a pair of hole spins. In this Letter we present a theoretical model, which gives a consistent explanation to the anomalous magnetic properties of this material.   


\end{abstract}
\maketitle

The single layer cuprate, HgBa$_2$CuO$_{4+\delta}$~(Hg1201), has not only the most symmetric crystal structure and the highest transition temperature, with a maximum value of 97K, but it also displays marked deviations in key properties from other single layer cuprate superconductors. NMR experiments on underdoped Hg1201 observe the usual charge disorder, but the standard magnetic disorder is not seen~\cite{Itoh1998,Rybicki2009,Haase2012,Rybicki2012}. In addition, neutron scattering finds two finite energy local excitations not observed elsewhere~\cite{Li2010,Li2012}. Li~\textit{et. al.}~\cite{Li2010,Li2012} interpreted these two triplet modes as a fingerprint of circulating orbital currents within the CuO$_6$ octahedra. However, this explanation was not supported by recent NMR measurement on Hg1201~\cite{Mounce2013}, finding no static magnetic fields that would accompany circulating orbital currents. 

NMR experiments also show an absence of antiferromagnetic~(AF) order in Hg1201 down to the lowest hole densities estimated at $\sim 5\%$~\cite{Itoh1998}. This value is well below $10\%$ predicted for the AF quantum critical point~(QCP) in an ideal underdoped CuO$_2$ layers by both theory~\cite{himeda1999} and experimental studies of multilayer Hg-cuprates~\cite{Mukuda2012, Chen2012}. 
Finally, although the O atom is strongly electronegative and generally favors an ionic state, O$^{2-}$, a substantial fraction of the holes is not inserted into the CuO$_2$ planes and remain on the dopant site~(O$^d$)~\cite{Jorgensen1997,Antipov2002}. Understanding the origin of these special features in the most ideal cuprate is a highly relevant challenge to theory. In this letter we take a closer look at the actual crystal structure of Hg1201 and propose a consistent explanation of all these anomalies.

In Hg1201 the O-dopants enter the Hg-layer, which is symmetrically placed halfway between neighboring CuO$_2$ layers. As a result the the loacal enhancement of the hole density will be equal in the layers above and below O$^d$. This structure is clearly different from other single layer cuprates, e.g. La$_{2-x} $Sr$_x$ CuO$_4$, where the dopant injects a single hole into a single nearby layer. A localized single hole in a CuO$_2$ layer generally is accompanied by a free spin. RKKY-like coupling causes the magnetic disorder at the nuclei observed in NMR experiments~\cite{Itoh1998}.

The more complex structure of the dopant O ion in Hg1201 requires a more detailed investigation of the charge distribution and accompanying spin distribution around the dopant site. Diffuse neutron scattering experiments by Jorgensen and coworkers \cite{Jorgensen1997} reported two different sites for the O dopant (O$^d$) in the Hg layer. At higher densities O$^d$ is predominantly at the O(3) site, which is symmetrically placed at the center of a square of Hg ions. But at lower densities, O$^d$  chooses predominantly an asymmetric site, O(4), lying close to a single Hg ion. In this case a single strong Hg-O(4) bond of $\sim 2$\AA~in length forms, whereas the larger Hg-O separation from the O(3) site greatly weakens hybridization between the planar 2p states of the O ions and the neighboring Hg ions. 
  
\begin{figure}[h]
  \begin{center}
	\includegraphics[width=1.\columnwidth, bb = 0 0 378 327]{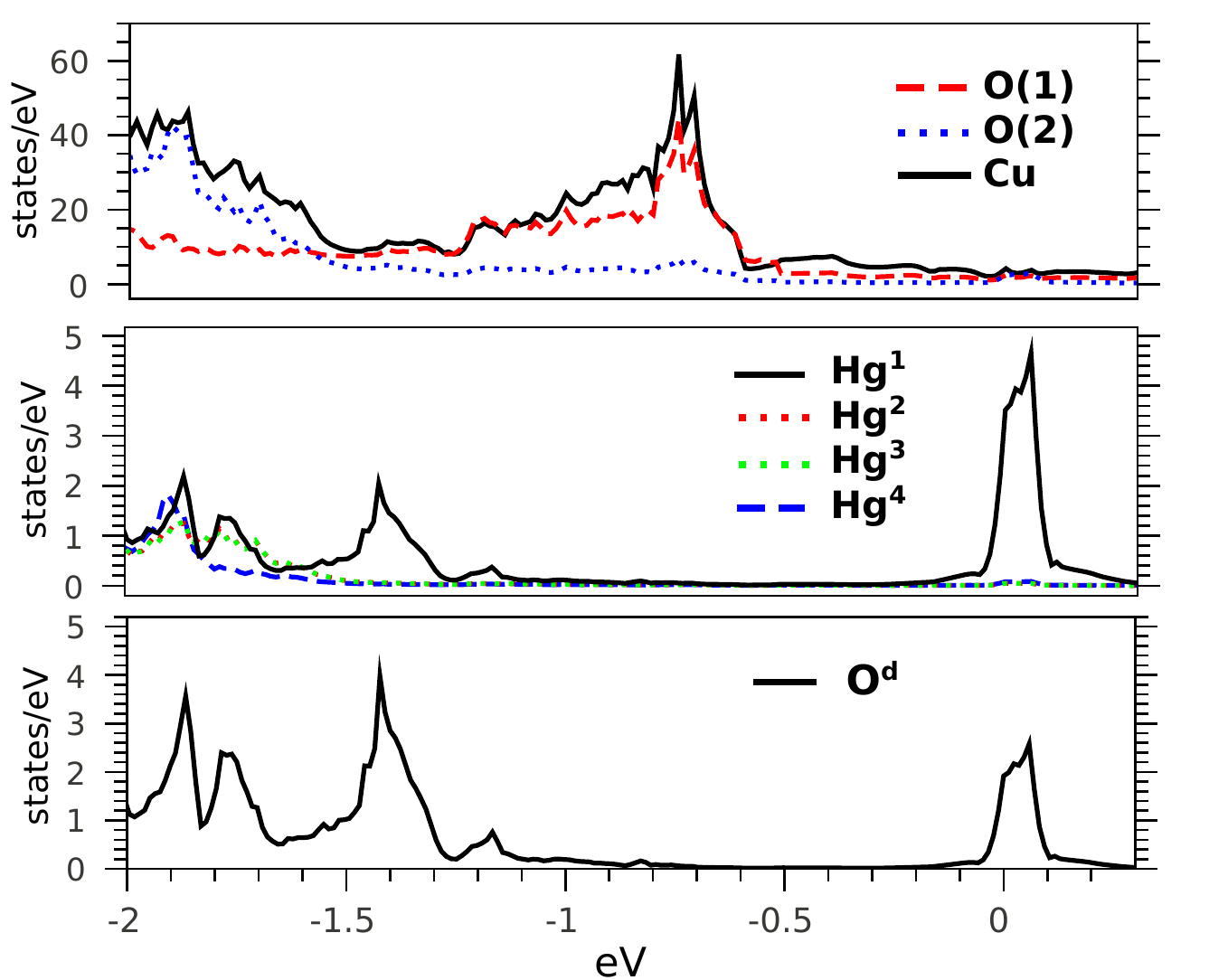}
  \end{center}
  \caption{Projected DOS for $\delta$=$0.125$ O(4) doping. Hg$^{1}$ (Hg$^4$) is the nearest (furthest) Hg-ions to the O(4) impurity in the central cell. Note the different scales in all three panels.}
  \label{fig:LDA}
\end{figure}

The electronic structure of the O$^d$ can be investigated using first-principles LDA calculations for a periodic supercell containing a single O$^d$. Supercell LDA calculations for the symmetric O(3) site were carried out by Ambrosch-Draxl et al.\cite{Ambrosch-Draxl2003}, found an essentially  ionic structure for this weakly hybridized O$^d$ but one with only partial occupancy of the uppermost O-valence states. These states overlap in energy with the uppermost CuO$_2$ layer states leading to a hole density which is shared between the O$^d$ and the neighboring CuO$_2$ layers. As remarked above, a reduced hole concentration in the layers is found experimentally~\cite{Jorgensen1997,Antipov2002}. In the case of the O(4) site, we have performed LDA calculations~\footnote{Supercell calculations were perfomed in VASP~\cite{VASP} {\it ab initio} package using PAW method~\cite{VASP, VASP-PAW}. For the DOS calculations illustrated in Fig.~\ref{fig:LDA} a $\Gamma$-centered $k$-point mesh of $10\times10\times8$ and experimental lattice constants~\cite{PRB} were used.} for a periodic 8-unit supercell containing one occupied O(4) site. The local density of states (see Fig. 1), has a filled bonding state below a partially filled anti-bonding state, indicating that O(4) also has holes distributed between the O$^d$ site in the Hg plane and sites in the nearby CuO$_2$ planes, again in agreement with experiment.

  We now turn to the analysis of the spin states and magnetic properties resultant from these two possible positions of O$^d$. For this purpose we examine how the spins of the holes moving in the adjacent CuO$_2$ planes are coupled via superexchange paths through the O$^d$. The suppressed magnetic inhomogeneity suggests that the two holes form a singlet ground state. In this case the two weakly-dispersive magnetic modes in Hg1201 observed with neutron scattering~\cite{Li2010,Li2012} can be attributed to the triplet excited states above such a ground state. The analysis of spin correlations of dopant holes localised around O$^d$ that we present below, supports such a scenario especially for the O(4) dopant location.

We start our analysis by considering electrons moving in the CuO$_2$ planes. These are described by a 3-band Hubbard model containing the O(1)-2$p_x$, 2$p_y$ and Cu-3$d_{x^2-y^2}$ orbitals. Early on it was shown that this description can be reduced to a 1-band Hubbard model~\cite{Zhang1988,Hybertsen1990} 
\begin{equation}
H_{0}=-t\sum_{nn}(c^{\dagger}_i c_j + h. c.) -t'\sum_{nnn}(c^{\dagger}_i c_j + h. c.) + \sum_i U n_{i\uparrow} n_{i\downarrow}
\label{ham0}
\end{equation}
Typical parameters for a  CuO$_2$ plane were estimated by Hybertsen et. al~\cite{Hybertsen1990}, who found values for nearest~(nn) and next-nearest~(nnn) neighbor hoppings, $t$ and $t'$ respectively, and the onsite Hubbard repulsion $U$ to be 
\begin{eqnarray}
  t=0.43~\text{eV},\quad t'/t=-0.16,\quad U/t=12.6
\end{eqnarray}
Note, however, that if we assume the attractive potential of the O$^d$ dopant confines the hole motion to a Cu$_4$ plaquette, there are just 4 sites and the kinetic energy is reduced. If we still take $U/t=12.6$, we are too close to the infinite $U$ limit where a Nagaoka effect applies, leading to a high spin S=3/2 groundstate\cite{Nagaoka1966} for 3 electrons on a Cu$_4$ plaquette .  However, a more physical low spin value can be obtained by using a smaller value, $U/t=6.0$ giving a ground state with $S= 1/2$. This is the simplest model to describe both charge and magnetic inhomogeneities similar to that of a hole localized in the CuO$_2$ plane around a Sr$^{2+}$ dopant in La$_{2x}$Sr$_x$CuO$_4$~\cite{Itoh1997,Itoh1998}.

Next we model the superexchange path coupling the spins of a pair of holes in Cu$_4$ plaquettes above and below a O$^d$ dopant. The hopping of the holes from the planes to the O$^{d}$ site leads to a superexchange interaction. The smallest local model to describe this behavior is a Cu$_4$O$_4$-O$^d$-Cu$_4$O$_4$ cluster using a 1-band Hubbard model to describe each Cu$_4$O$_4$ plaquette. The O(4) dopant state is then approximated by solving a small cluster containing the 2$\times$4-plaquette states and the antibonding Hg-O state, to give a total of 9 states with 8 electrons. To represent the symmetric O(3) dopant we include the pair of orthogonal active O-$2p$, making a cluster with 10 states and 10 electrons. In each case the cluster is small enough that it can be easily diagonalised.
Our aim is put forward a qualitative explanation of the special properties of Hg1201 listed above, rather than to attempt a numerically accurate description, which would require larger clusters. 

\begin{figure}[h]
  \begin{center}
	\includegraphics[width=0.5\columnwidth, bb=0 0 337 457]{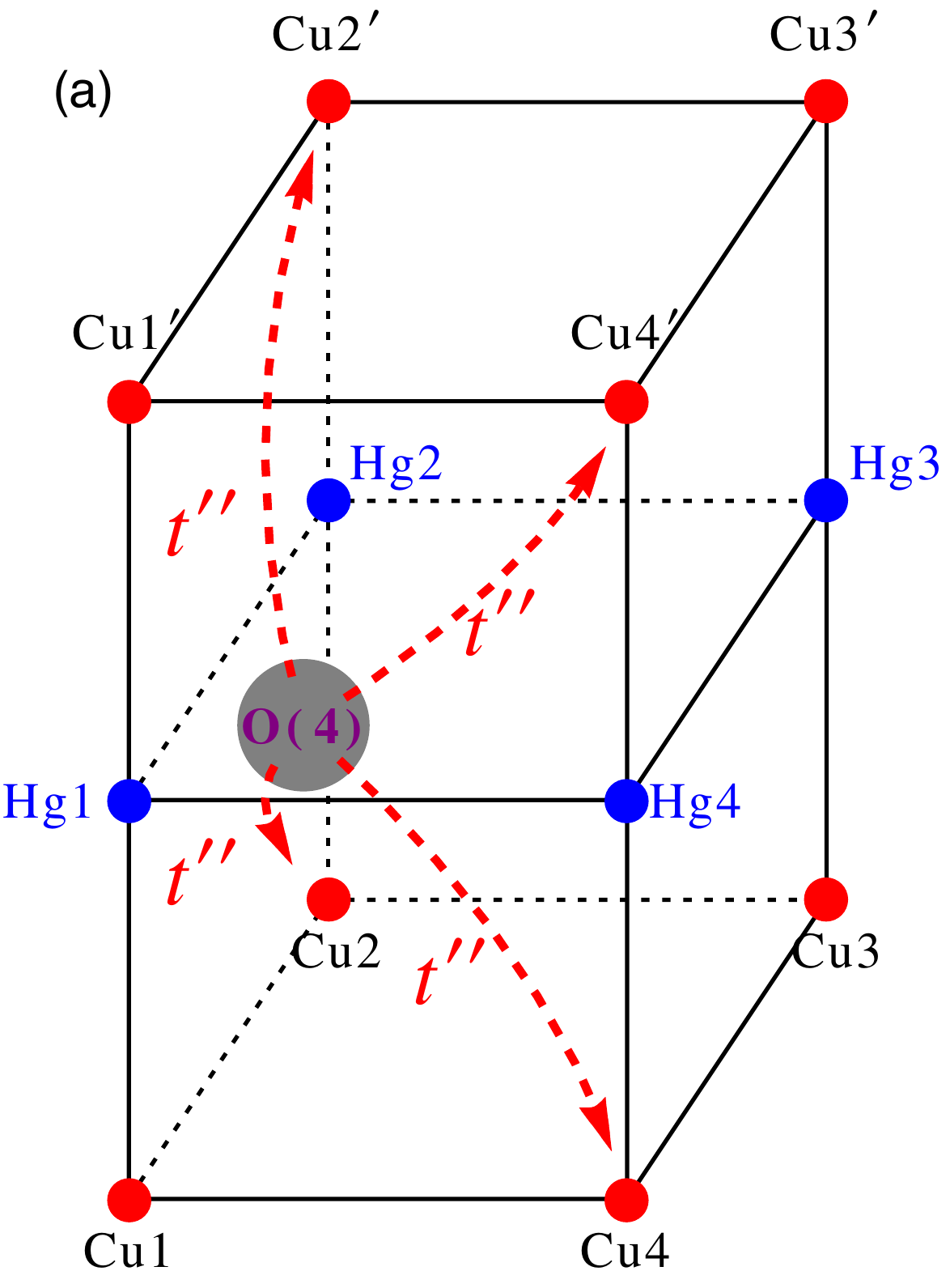}\includegraphics[width=0.5\columnwidth, bb=0 0 337 457]{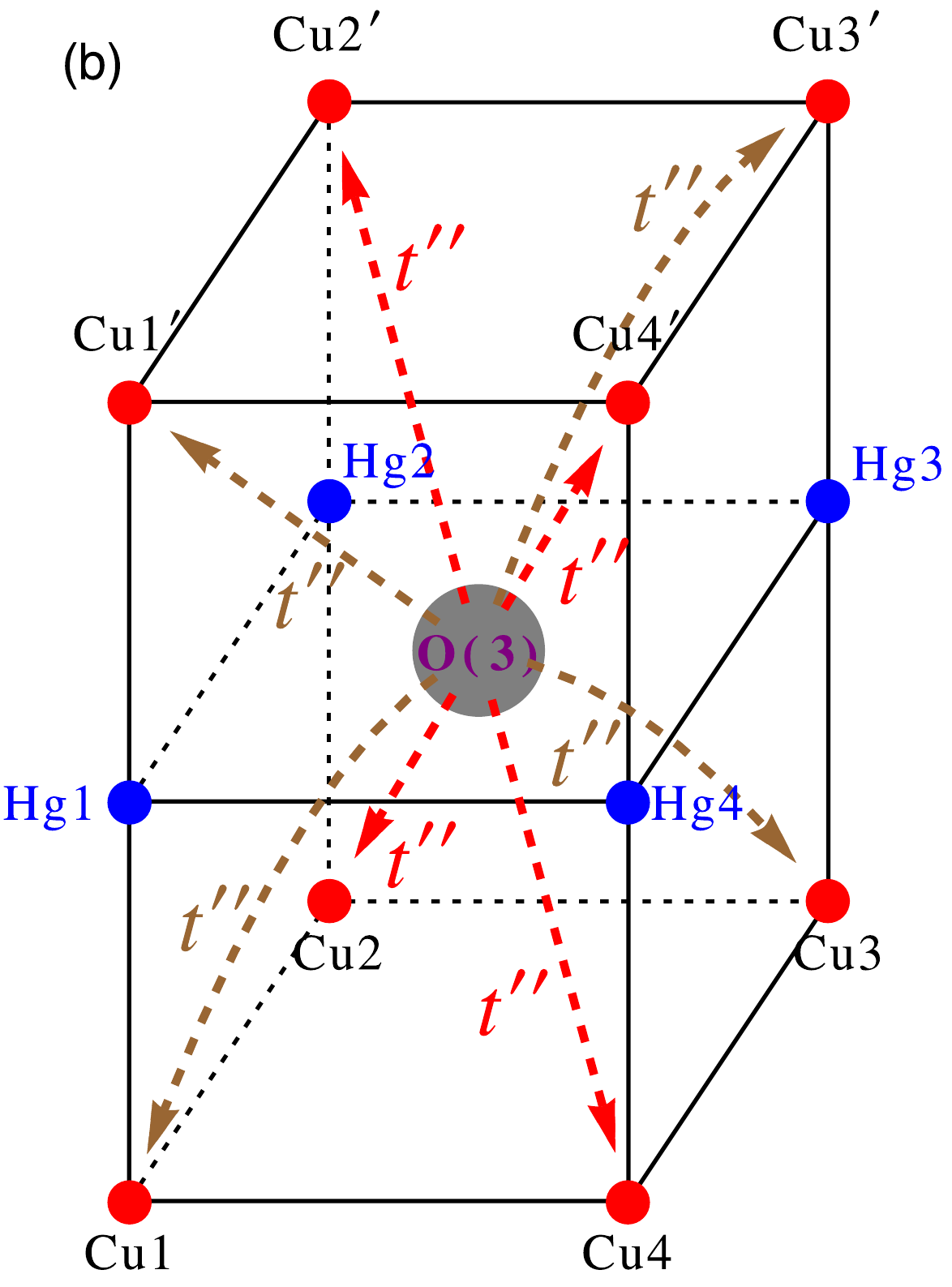}
  \end{center}
  \caption{The superexchange paths for holes from oxygen defect onto the planar Cu sites: (a) O(4): Holes in the Hg plane enter the antibonding Hg1-O(4) state (b) O(3): Holes enter the 2p-O(3) states weakly hybridised with (Hg1-Hg3) and (Hg2-Hg4)}
  \label{fig:CuO}
\end{figure}

 We begin with the O$^{d}$ sitting on the O(4) site, displaced towards Hg1 along the diagonal as shown in Fig. \ref{fig:CuO} (a).  It forms a Hg1-O(4) bond and only the anti-bonding state is relevant. An electron can hop from this anti-bonding state to the O(2) ions (apical oxygen) below and above the Hg plane. Since $p_z$ orbital on O(2) is orthogonal to the 3$d_{x^2-y^2}$ on nn Cu1 and Cu$1'$, an electron cannot directly hop onto these sites, but it can hop onto their higher energy 4$s$ orbital and then to 2$p_x$ and 2$p_y$ on the neighboring O(1) planar O-ions. The superexchange paths from O(4) defect to the planar Cu$_4$ plaquettes are  O(4)$\xrightarrow{\text{Hg1}}$(Cu2, Cu4, Cu$2'$, Cu$4'$) with
\begin{eqnarray}
  H_{\text{O(4)-Cu}}=-t''\sum_{\text{O}\rightarrow i}(c_{i\sigma}^\dag c_{\text{O}\sigma}+\text{h.c.}),
  \label{eq:o4cu}
\end{eqnarray}
where $c_{\text{O}\sigma}$ is the electron operator on O(4) defect site and $\text{O}\rightarrow i$ denotes the paths O(4)$\xrightarrow{\text{Hg1}}$(Cu2, Cu4, Cu$2'$, Cu$4'$) labeled with dashed red arrows in Fig. \ref{fig:CuO} (a).  The potential and interaction terms on O(4) defect are written as
\begin{eqnarray}
  H_{\text{O4}}=\epsilon_p\sum_{\sigma}c_{O\sigma}^\dag c_{O\sigma}+U_p n_{O\uparrow}n_{O\downarrow},
  \label{eq:o4}
\end{eqnarray}
where $\epsilon_p$ stands for the energy of the antibonding state.
We fix the hopping amplitude into the anti bonding Hg-O state and its on-site repulsive interaction as
\begin{eqnarray}
  t''/t=0.5,\quad U_p/t=1,
\end{eqnarray}
and tune the oxygen energy position $\epsilon_p$ to control the hole occupation in the CuO$_2$ plane. The total Hamiltonian for the cluster is now obtained by combining the Hubbard term Eq.~(\ref{ham0}) with the dopant-associated terms Eqs.~(\ref{eq:o4cu}-\ref{eq:o4}). 

\begin{figure}[t]
  \begin{center}
	\includegraphics[width=\columnwidth, bb = 16 18 422 343]{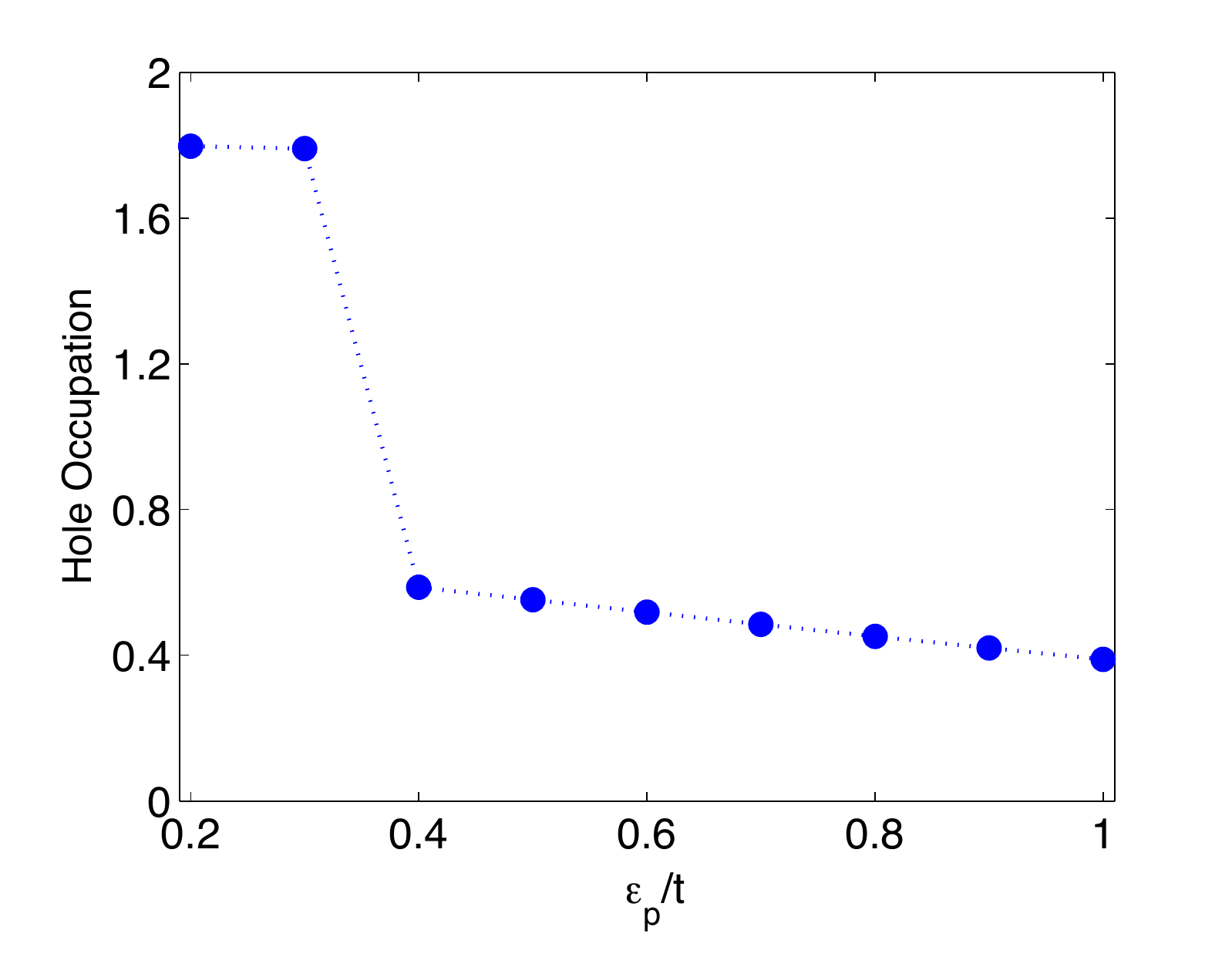}
  \end{center}
 \caption{Occupation of holes in the CuO$_2$ plane for the O(4) defect in Hg1201 as a function of the antibonding state energy $\epsilon_p$.}
  \label{fig:occ}
\end{figure}

After diagonalizing the Hamiltonian, for small values of $\epsilon_p$ the O(4) site is found to be almost fully filled by electrons and two holes are almost completely injected into the nearby Cu$_{4}$ plaquettes as illustrated in Fig.~\ref{fig:occ}. The system can be approximated as two weakly coupled $S=1/2$ states, resulting in nearly degenerate singlet and triplet states. In this case, the dopant induced charge and magnetic inhomogeneities are similar. With increasing $\epsilon_p$, the occupation of holes in the CuO$_2$ planes decreases and only a smaller fraction of the holes are injected into the planar sites, as found by Jorgensen and collaborators\cite{Jorgensen1997}. Such behavior is also confirmed in our LDA calculations (see Fig.\ref{fig:LDA}). In this case the spin correlations of the two holes increase due to the stronger superexchange path via the O(4) defect.

\begin{figure}[b]
  \begin{center}
 \includegraphics[width=\columnwidth, bb = 4 -20 519 430]{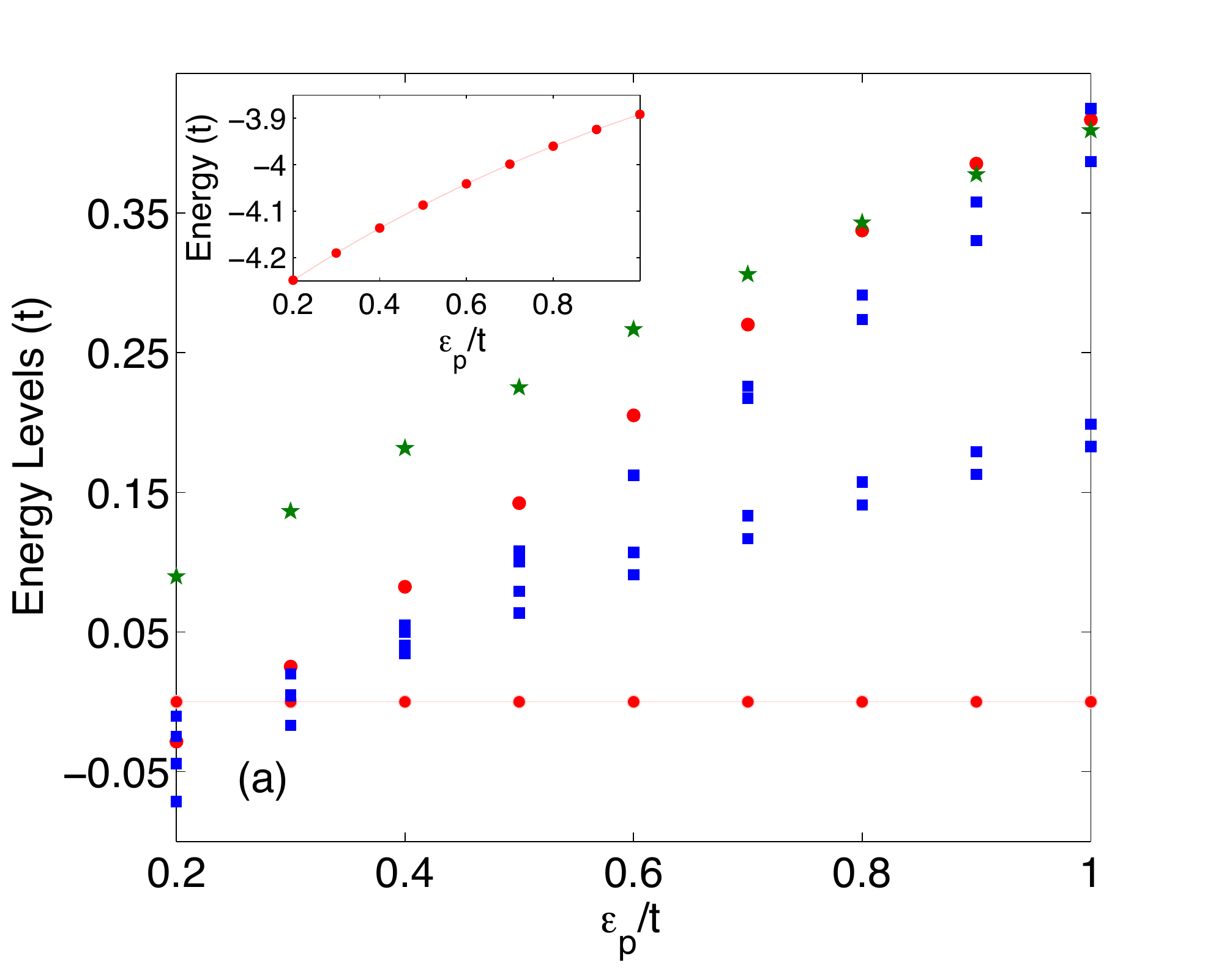}
 \includegraphics[width=\columnwidth, bb = 4 11 519 430]{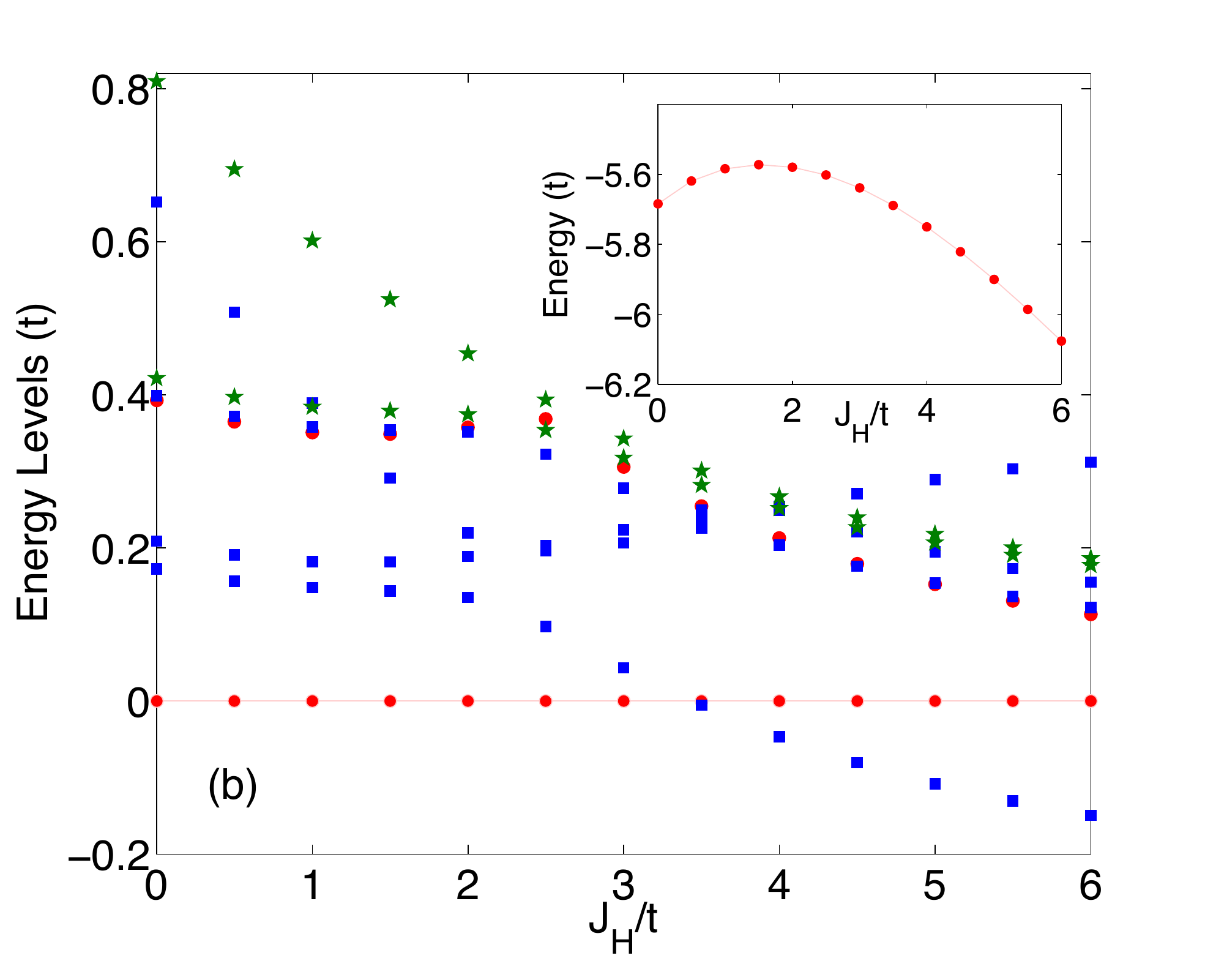}
 \end{center}
 \caption{(a) O(4) dopant site: $\epsilon_p$-dependent low energy levels for O(4) defects in the Cu$_4$-O-Cu$_4$ cluster. (b) O(3) dopant site: $J_H$-dependence of the energy levels for O(3) defect with the parameter set in Eq. (\ref{eq:paro3}). A solid red circle denotes a singlet state, a solid blue square a triplet and a solid green star a spin-$S=2$ state.  The energy of the reference singlet state is shown in the inset.}
  \label{fig:energy}
\end{figure}

The $\epsilon_p$-dependent occupation of holes in the twin Cu$_4$ plaquettes is shown in Fig. \ref{fig:occ}. The hole occupation dramatically changes around $\epsilon_p/t=0.3$ implying a level crossing in the ground state.  The $\epsilon_p$-dependent energy levels are shown in Fig.~\ref{fig:energy} (a).  When $\epsilon_p/t\geq0.4$, the singlet turns out to be the ground state and the lowest excited states are triplets. The two lower triplets start to separate from the higher energy levels. At $\epsilon_p/t=0.6$, our parameter set gives the hole occupation in the CuO$_2$ planes as 0.44 hole per oxygen defect and the two triplets have the energy 39 meV and 46 meV when $t=0.43$ eV. The two triplet modes differ in local spin correlations. The lower triplet has the spin correlation, $\langle \mathbf{S}_{\text{Cu2}}\cdot\mathbf{S}_{\text{Cu}2'}\rangle=-0.0354$, and the upper one has the correlation, $\langle \mathbf{S}_{\text{Cu2}}\cdot\mathbf{S}_{\text{Cu}2'}\rangle=0.0970$. 

As noted earlier, when doping level increases, the O(3) site occupation starts dominating in Hg1201~\cite{Jorgensen1997}. Since the  O(3) site is at the center of the Hg square, the main difference between the O(4) and O(3) positions is that the latter has two relevant $2p$ orbitals. In this case there are two different types of electron superexchange paths: O(3)$\xrightarrow{\text{(Hg1, Hg3)}}$(Cu2, Cu4, Cu$2'$, Cu$4'$) and O3$\xrightarrow{\text{(Hg2, Hg4)}}$(Cu1, Cu3, Cu$1'$, Cu$3'$) 
\begin{eqnarray}
  H_{\text{O3-Cu}}=-t''\sum_{\text{O}\rightarrow i}(c_{i\sigma}^\dag c_{\text{O1}\sigma}+c_{i\sigma}^\dag c_{\text{O2}\sigma}+\text{H.C.})
  \label{o31}
\end{eqnarray}
where $\text{O}\rightarrow i$ denotes O(3)$\xrightarrow{\text{(Hg1, Hg3)}}$(Cu2, Cu4, Cu$2'$, Cu$4'$) and O3$\xrightarrow{\text{(Hg2, Hg4)}}$(Cu1, Cu3, Cu$1'$, Cu$3'$) labeled as the red and brown dashed arrows, respectively, in Fig. \ref{fig:CuO} (b). $c_{\text{O1,2}}$ are the electron operators of the two states on O(3) defects.

Since there are two relevant orthogonal states, Hund's rule interaction is relevant on the O(3) defect
\begin{eqnarray}
  H_{\text{O3}}&=&\epsilon_{p}\sum_{\sigma}(c_{\text{O1}\sigma}^\dag c_{\text{O1}\sigma}+c_{\text{O2}p_y\sigma}^\dag c_{\text{O2}\sigma})\nonumber\\
  &+&U_{p}(n_{\text{O1}\uparrow}n_{\text{O1}\downarrow}+n_{\text{O2}\uparrow}n_{\text{O2}\downarrow})
  +U'_{p}n_{\text{O1}}n_{\text{O2}}\nonumber\\
  &-&J_H\mathbf{S}_{\text{O1}}\cdot\mathbf{S}_{\text{O2}}
  \label{o32}
\end{eqnarray}
where $U_p$ and $U'_p$ are for the intraband and interband interaction and $J_H$ is the Hund's rule interaction. We fix the parameters as
\begin{eqnarray}\label{eq:paro3}
  t''/t=0.5,~ U_p/t=1,~ U_{p'}/t=1,~ \epsilon_p/t=-1,
\end{eqnarray}
and tune the Hund's rule interaction $J_H$. The total Hamiltonian for the O(3)-dopant is given by a sum of $H_0$ of Eq.~(\ref{ham0}) and the two terms Eqs.~(\ref{o31}-\ref{o32}). 

The $J_H$-dependent energy spectrum for holes around the O(3) defect is shown in Fig.~\ref{fig:energy} (b). For a small value of $J_H$ the ground state is a singlet. With increasing $J_H/t$, the triplet state decreases in energy and eventually becomes a ground state at $J_H/t\geq3.5$. At $J_H/t=4$, there are 1.87 holes on the O(3) defect in the triplet ground state. The spin correlation is $\langle\mathbf{S}_{\text{O1}}\cdot\mathbf{S}_{\text{O2}}\rangle=0.208$ implying the parallel alignment of the spins of holes on O(3) defect as expected according to the standard Goodenough-Kanamori rule. Note, LDA calculations for the O(3) site in Hg1201~\cite{Ambrosch-Draxl2003} and for the same site in the 3-layer Hg1223~\cite{Singh1994} give a substantial fraction of the doped holes on the site at strong overdoping.

 We conclude that in Hg1201 the position of the divalent O$^d$ site, situated in the Hg-layer halfway between two CuO$_2$ layers, strongly changes the magnetic and spin properties relative to the standard case of a singly charged dopant next to a single CuO$_2$  layer. Our analysis shows that at underdoping, when the oxygen defect is mainly at the asymmetric O(4) site, the superexchange interaction between dopant holes through the O(4) defect leads to a singlet ground state. This gives a natural explanation of the striking difference of charge and magnetic inhomogeneities in the NMR experiments in Hg1201~\cite{Itoh1998,Rybicki2009,Haase2012,Rybicki2012}. Above the singlet ground state, there are two triplet excited states, which can be characterized by the different local spin correlations between the nearby CuO$_2$ planes. The numerical result from the small cluster however underestimates the non-local effects such as the effect of finite hole concentrations in the CuO$_2$ planes. With the parameter values we have chosen, there are two triplet modes with excitation energies 39 meV and 46 meV. These energies are of the same order as the values measured in the neutron scattering experiments~\cite{Li2010,Li2012}.  

Lastly, we note that the weak spin disorder and high symmetry in Hg1201 makes it a specially interesting cuprate for neutron investigation of low energy spectrum of the magnetic excitations in the underdoped pseudogap state. In particular, the low energy spectrum should not be contaminated by the low energy free local spins associated with localized single planar holes. The NMR experiments also did not find any antiferromagnetic order down to a doping of $\sim 5\%$. This doping is well below the QCP for the onset of coexisting superconducting and AF order in an ideal single layer, deduced from NMR experiments on the multilayer Hg-cuprates~\cite{Mukuda2012,Chen2012,himeda1999}. It is worth remarking that the suppression of the AF QCP to very low hole densities (less than $5\%$) in Hg1201 does not weaken superconductivity, which shows a T$_c$ = 50K at $5\%$ density~\cite{Itoh1998}. This observation would seem to question proposals that proximity to the AF QCP acts to stabilise superconductivity. Rather it favors singlet pair correlations as a key factor.

We used the ALPS~\cite{bauer2011alps} to carry out the sparse diagonalizations in this paper~\footnote{~https://alps.comp-phys.org}. We are grateful to L. Wang for his help with the ALPS package, and to N. Spaldin for giving access to the VASP license. We also thank M. Greven and M. Sigrist for useful discussions. The work in Switzerland was supported by the Swiss Nationalfonds.  
\bibliography{Revised26Nov-copy}

\end{document}